\documentstyle[aps,preprint,prd,12pt]{revtex}
\begin{document}
\def\be{\begin{eqnarray}}
\def\ee{\end{eqnarray}}
\def\heli{\Upsilon}
\def\densi{\varrho_s}
\def\coup{K}
\renewcommand{\baselinestretch}{1.}
\baselineskip 3ex
\begin{titlepage}
\begin{flushright}
KL-TH-93/10\par
CERN-TH.6885/93
\end{flushright}
\vspace{1ex}
\begin{center}
{\LARGE Critical Behaviour of the 3{$\!D$ $XY$}-Model:}\\
\vspace{0.3 cm}
{\LARGE                   A Monte Carlo Study}

\vspace{0.75 cm}
{\large    Aloysius P. Gottlob}

\vspace{0.2 cm}
{\it    Universit\"at Kaiserslautern, D-6750 Kaiserslautern, Germany}
\vspace{0.35 cm}

and

\vspace{0.35 cm}
{\large     Martin Hasenbusch}

\vspace{0.2 cm}
{\it    CERN, Theory Division, CH-1211 Gen\`{e}ve 23, Switzerland}
\end{center}
\vspace{2cm}
\setcounter{page}{0}
\thispagestyle{empty}
\begin{abstract}\normalsize
We present the results of a study of the three-dimensional $XY$-model
on a simple cubic lattice using the  single cluster updating  algorithm
combined  with improved estimators. We have measured the susceptibility and
the  correlation length for various couplings in the high temperature phase
on lattices of size up to $L=112$. At the transition temperature we studied
the fourth-order cumulant and other cumulant-like quantities
on lattices of size up
to $L=64$. From our numerical data we obtain for the critical coupling
$\coup_c=0.45420(2)$, and for the static critical exponents
$\gamma /\nu=1.976(6)$ and  $\nu=0.662(7)$.
\end{abstract}
\nopagebreak
\begin{flushleft}
\vspace{3 cm}
KL-TH-93/10\newline
CERN-TH.6885/93\\
May 1993
\end{flushleft}
\vspace{1ex}
\end{titlepage}

\section{Introduction}
\noindent
Three-dimensional classical $O(N)$ vector models are of great interest, both as
simplest statistical models with a continuous symmetry and as a lattice version
of the scalar quantum field theory with $(\vec{\phi^2})^2$-interaction.
In particular, the 3$D$ $O(2)$ model, also called the
$XY$-model, is relevant to the
critical behaviour of a number of physical systems, such as the phase
transition of superfluid $^4$He and magnetic systems  with planar spin
Hamiltonians.
Quantitative knowledge of the critical behaviour of the $O(N)$ vector models
is mostly based on the field theoretic renormalization group techniques at
dimension $D=3$ \cite{guillou80} and the $\epsilon$-expansion
\cite{wilson72,guillou85a}. Very accurate values of the
critical exponents are among the most successful predictions of
these approaches. In addition, the analysis of high
 temperature series expansions \cite{dombgreen}
 provides estimates for the critical temperature
 of particular lattice models.

Monte Carlo simulations  have also
succeeded in providing detailed information about the critical behaviour of
the 3$D$ $O(N)$ vector models, but
only in the case of the three dimensional Ising model $(N=1)$
\cite{ferrenberg91a,baillie92a} is an accuracy close to
that of analytic calculations reached.

The difficulty encountered in Monte Carlo simulations with local updates is
the critical slowing down near a phase transition. Considerable progress has
been achieved during the last 6 years with the development of efficient
non-local Monte Carlo algorithms which overcome critical slowing down to a
large extent \cite{swend+sokal}.

In the present paper we extend previous Monte Carlo studies of the $3D$ $XY$
model \cite{we90a,janke90a} where cluster algorithms \cite{Wang,wolff89a} were
first applied to simulate this model. These studies were performed
on vector computers, using a moderate amount of CPU-time. Since optimal
vectorization cannot be reached for cluster algorithms, we used modern
RISC stations for the present study. Using about two months of CPU-time we
were able to simulate larger lattices and reached a considerably better
statistical accuracy than in the previous studies. This accuracy allowed us to
control  systematic errors in the estimates for  the critical
coupling and the critical exponents.

This paper is organized as follows.
In section 2 we give the definition of the model and
describe the cluster updating algorithm,
in section 3 we give our results for the high temperature
phase, while section 4 contains our data obtained in the critical region.
In section 5 we compare our results with those of previous studies.

\section{Cluster Update Monte Carlo of the
3{\protect\boldmath$D$ $XY$}-Model}
We study the $XY$-model in three dimensions defined by the
partition function
   \begin{equation}
   {\cal Z}= \prod_{i\in\Lambda}\int_{S_{1}}\!\!ds_i\;
   \exp(\coup\sum_{\langle i,j\rangle}\vec{s}_i%
   \cdot\vec{s}_j)\; ,
   \label{a}
   \end{equation}
where $ \vec{s}_i$ is a two dimensional unit vector, the summation is
taken over all nearest neighbour pairs of sites $i$ and $j$  on a
simple cubic lattice $\Lambda$ and $\coup=\frac{J}{k_b T}$ is the coupling,
or more precisely, the reduced inverse temperature.

For ferromagnetic interactions $J>0$, the $XY$-model has a second-order phase
transition separating a low temperature phase with non-zero magnetization from
a
massive disordered phase at high temperatures. This phase transition can be
viewed alternatively as due to Bose condensation of spin waves
\cite{frohlich76} or the unbinding of vortex strings
\cite{frohlich82,kohring86}.

A major difficulty encountered in Monte Carlo simulation at second-order phase
transitions is critical slowing down. The
autocorrelation time $\tau$, which is roughly the time needed to generate
statistical independent configurations, grows as $\tau \propto L^z$ at
criticality, where $L$ is the linear size of the system and
$z$ is the dynamical critical expontent. Random walk
arguments indicate that local updates like the Metropolis algorithm result in
$z=2$, which is consistent with the numerical finding for the $3D$ $XY$-model
\cite{we90a}.

In the case of $O(N)$ vector models, critical slowing down can be drastically
reduced, using cluster algorithms \cite{we90a,wolff89a,janke90a}.

In the present work we employ the single cluster algorithm which was introduced
by Wolff \cite{wolff89a}. Let us shortly recall the steps of the update.
First choose randomly a reflection axis in the $I\!\!R^2$ plane. Denote the
component of the spin $\vec{s}_i $ which is parallel to this reflection axis
by $s_i^{\|}$ and that which is orthogonal by $s_i^{\bot}$.
Then choose randomly a site $i$ of the lattice as a starting point for the
cluster $\cal C$. Visit all neighbour sites $j$  of $i$. These sites join the
cluster with the probability
  \be
  p(i,j) & = & 1 - \exp(-\coup (s_i^{\bot} s_j^{\bot} + | s_i^{\bot} s_j^{\bot}
  | )).
  \ee
After this is done, visit the neighbours of the new sites in the cluster and
add them to the cluster with probability $p(i,j)$ which is given above. Iterate
this step until no new sites enter the cluster. Now flip the sign of all
$s^{\bot}$ contained in the cluster.

\section{Numerical results in the high temperature phase.}
\subsection{Observables to be measured}
Let us first summarize the definitions of the observables that we studied.
The energy density
is given by  the two-point correlation function $G(x_i,x_j) =
\langle \vec{s}_i\vec{s}_j\rangle$  at distance one
  \be
  E =  \frac{1}{3 L^3} \sum_{\langle i,j \rangle}\langle \vec{s}_i
  \vec{s}_j \rangle .
  \ee
The specific heat of the system at constant external field is defined by the
derivative of the energy density with respect to  the inverse temperature.
It can be obtained from the fluctuations of the energy $H  = -
   \sum_{\langle i,j \rangle}   \vec{s}_i
  \vec{s}_j $
  \be
  C_h = \frac{1}{L^3} \left( \langle H^2 \rangle - \langle H \rangle^2 \right).
  \ee
The magnetic susceptibility $\chi$ gives the reaction of the magnetization
$m = \sum_{i\in \Lambda} \vec{s}_i$ to an external field. In the high
temperature phase one gets
  \be
  \chi = \frac{1}{L^3} \langle   m^2 \rangle  ,
  \ee
since $ \langle m \rangle = 0$.

Cluster algorithms enable one to reduce the variance of the expectation values
in the high temperature phase by using improved estimators
\cite{wolff90a,hasenbusch90a}. The improved estimator of the  magnetic
susceptibility is given by
  \be
  \chi_{imp} &=& \langle \; \frac{2}{|{\cal C}|}\; ( \sum_{i \in {\cal C}}
  s_i^\bot )^2\;\rangle,
  \ee
where $|\cal C|$ denotes the number of spins in the cluster $\cal C$.

There are two common definitions of a correlation length $\xi$. The exponential
correlation length $\xi_{exp}$ is defined via the decay of the two-point
correlation function at large distances
  \be
  \xi_{exp}&=&\lim_{|x_i -x_j|\to \infty}\frac{-|x_i - x_j|}{\log G(x_i,x_j)},
  \ee
which is equal to the inverse mass gap.
For the measurement of the exponential correlation length we consider the
correlation function
  \be
  \overline{G}(t) \equiv \langle O_0 O_t \rangle \propto   \;
  \left( \exp(\frac{-t}{\xi_{exp}}) +  \exp(\frac{-(L-t)}{\xi_{exp}})\right),
  \label{corfunc}
  \ee
of the translational invariant time slice magnetization
$O_t = \sum_i \vec{s}(x_{i},t)$.

The second-moment correlation length is defined by
  \be
  \xi_{2^{nd}}=\left(\frac{(\chi/F)-1}{4 \sin^2(\pi/L)} \right)^{\frac{1}{2}},
  \label{xi2def}
  \ee
with $F=\widehat{G}(k)|_{|k|=2\pi/L}$, where $\widehat{G}(k) = \sum_{j\in
\Lambda} \langle \exp(ikx_j)  \vec{s}_0\vec{s}_j \rangle$ is the Fourier
transform of the two-point correlation function  and $\chi$ the magnetic
susceptibility. For more details see for example ref. \cite{sokal92a}.
The two definitions of the
correlation length do not coincide, since in $\xi_{exp}$ only the first
excited state enters, while in the case of $\xi_{2^{nd}}$ a mixture of the
full spectrum is taken into account. However, near the critical point the two
quantities should scale in the same way.
As for the magnetic susceptibility there exist improved estimators for the
two definitions of the correlation length. The improved estimator of the
two-point correlation function is given by
  \be
  \langle\vec{s}_i \vec{s}_j \rangle_{imp} & = &
  \left\langle   \frac{2}{|{\cal C}|} \; \delta_{ij} ({\cal C})\; s_i^\bot
  s_j^\bot \right\rangle,
  \ee
where $\delta_{ij}({\cal C})=1$ if $i$ and $j$ belong to the same cluster
$\cal C$, otherwise $\delta_{ij}({\cal C})=0$ \cite{wolff90a,hasenbusch90a}.
For $\xi_{2^{nd}}$ one has to provide a $F_{imp}$. This is given by the
Fourier transform of the improved two-point correlation function
\cite{hasenbusch90a,janke92a}
  \be
  F_{imp} = \left.\widehat{G}(k)_{imp}\right|_{|k|=\frac{2 \pi}{L}} ,
  \ee
with
  \be
  \widehat{G}(k)_{imp} = \left\langle
  \frac{2}{|{\cal C}|}
  \left( \left( \sum_{i \in {\cal C}} s^\bot_i \cos(k x_i)\right)^2 +
  \left( \sum_{i \in {\cal C}} s^\bot_i \sin(k x_i)\right)^2 \right)
  \right\rangle .
  \ee
The helicity modulus describes the reaction of the
system to a suitable phase twisting field \cite{fisher73a}.
The lattice definition of the helicity modulus is given
by

  \be
  \Upsilon_\mu = \frac{1}{L^3}
  \left< \sum_{\langle i,j \rangle}
  s_i s_j (\epsilon_{\langle i,j \rangle} \mu)^2\right >
  - \frac{\coup}{L^3} \left < (\sum_{\langle i,j \rangle}
  (s_i^1 s_j^2 + s_i^2s_j^1)\epsilon_{\langle i,j\rangle} \mu)^2\right >,
  \label{helidef}
  \ee
where $\mu$ is a unit vector in $x,y$ or $z$ direction and
$\epsilon_{\langle i,j \rangle}$ the unit vector connecting  the sites $i$ and
$j$ \cite{liteitel}.

\subsection{Monte Carlo Simulations}
In order to obtain an estimate of the critical coupling $K_c$ and
determine static critical exponents, we
have done 15 simulations at couplings from $K =  0.4$ up to $K=0.452$ on
lattices of linear size $L=24$  up to $L=112$. The simulation parameter and the
results of the runs are given in Tables\ \ref{htresult} and
\ref{corrlength}. The statistics is given in terms of $N$ measurements taken
every $N_0$ update steps.  $N_0$ is chosen such that approximately $N_0
\times \langle {\cal C} \rangle = L^3$ and hence the whole lattice is updated
once for a measurement. We estimated the statistical errors  $\sigma_A$ of
expectation values $\langle A \rangle$ from
  \be
  {\sigma_A}^2 =  \frac{\langle A^2 \rangle - \langle A \rangle ^2}{N/(2 \tau)}
  \ee
and from a binning analysis. These error estimates were consistent throughout.
The statistical error of quantities which contain several expectation values we
calculated from Jackknife-blocking \cite{siam}.

\subsubsection{Finite-size effects}
We tried to avoid a finite-size scaling analysis. Hence we had to choose our
lattices  large enough to ensure
that deviations of the values of the observables from the
thermodynamic limit values are negligible.

We therefore have measured  the energy density $E$, the specific heat $C_h$,
the helicity modulus $\heli$ ,
the exponential correlation length $\xi_{exp}$ and the second-moment
correlation length $\xi_{2^{nd}}$  for fixed coupling $\coup = 0.435$ and
increasing system size $L=4$ up to $L=32$. The results are summarized
 in Table\ \ref{fsizetab}. The values of the observables
 obtained for $L=24$ and $L=32$ are
consistent within error bars. Furthermore, the values of the
helicity
modulus $\heli$ for $L=24$ and $L=32$ are consistent with $0$, which is the
thermodynamic limit value of the helicity modulus in
the high temperature phase. The correlation length at $K=0.435$ is
approximately
 $4$.
Hence we conclude, assuming scaling,
 that the systematical deviations from the thermodynamic limit are
smaller than our statistical errors for $L/\xi\ge 6$. This condition is
fulfilled  by all the simulation parameters of our runs given in Table\
\ref{htresult}.

\subsubsection{Energy density and specific heat}
The energy density $E$ shows, as expected, no singular behaviour close to the
critical temperature. In the scaling region the specific heat $C_h$ should
follow
  \be
  C_h  = C_{reg}  + C_0 \; \left(\frac{K_c - K}{K_c} \right)^{-\alpha},
  \ee
where $C_{reg}$ denotes the regular part of the specific heat and $\alpha$ is
the critical exponent of the specific heat. In order to  estimate $\alpha$ we
did a four-parameter least-square fit.
 However, it was not possible to extract
meaningful estimates. The best fit to the data leads to $K_c = 0.456(2)$,
$\alpha = 0.23(13)$ with $\chi^2/d.o.f. \approx 0.93$ and the relative errors
of the constants are about $100\%$. If we fix the critical coupling to
$K_c = 0.45420$ (this is our estimate obtained at criticality) the quality of
the fit gets worse. Therefore we assumed $\alpha = 0$ and fitted the data
following
  \be
  C_h  = C_{reg} + C_0 \; \log \left(\frac{K_c - K}{K_c} \right).
  \ee
The best three-parameter fit to our data leads to
$K_c = 0.4543(4)$, $C_{reg}=-0.49(20)$ and $C_0 = -1.61(7)$ with $\chi^2/d.o.f.
\approx 0.61$, where data with $\xi > 2.5$ are taken into account. This result
shows that our data for the specific heat, combined with extended ans\"atze,
are
compatible with an $\alpha=-0.007(6)$  obtained from the  hyperscaling relation
$\alpha= 2 - D \nu $  and the estimate $\nu=0.669(2)$
from resummed perturbation series \cite{guillou80},
but have no predictive power for the exponent $\alpha$.

\subsubsection{Magnetic susceptibility}
For comparison we give in Table\
 \ref{htresult} the results for the standard and
the improved susceptibilities. The statistical error of $\chi_{imp}$
is about 3.5 to 8 times smaller than the error of the
standard susceptibility. But one should remark that
the statistical error of the standard estimator
depends very much on how often one measures.
In the following we only discuss the results obtained with the improved
estimator. In order to estimate the critical
coupling and the susceptibility exponent $\gamma$   we performed
a  three-parameter least-square fit following the scaling law
  \be
  \chi =
  \chi_0\; \left(\frac{\coup_c-\coup}{\coup_c} \right)^{-\gamma}. \label{simk}
  \ee
We obtained $\gamma = 1.324(1)$, $\coup_c = 0.454170(7)$
and $\chi_0 = 1.009(2)$ with $\chi^2 /d.o.f. = 0.65$, when all data are
taken into account. In order to test the stability of the  results  we
successively discarded data points with small $K$. The results
of these fits are summarized in Table\ \ref{chikfit}. $\chi^2 /d.o.f.$ remains
small and the results for $\gamma$, $\coup_c$ and $\chi_0$ are consistent
within the error bars for all data-sets that we used.
But the small $\chi^2 / d.o.f.$ of the fits discussed above,
is of course, no proof for the absence of corrections to the scaling.
{F}rom renormalization group considerations \cite{wegner} one expects
confluent and analytical corrections of the type
  \be
  \chi(\coup) = \chi_0\;     \left(\frac{\coup_c -
  \coup}{\coup_c}       \right)^{-\gamma} + \chi_{conf.}\;
  \left(\frac{\coup_c - \coup}{\coup_c}
   \right)^{-\gamma+\Delta_1}+ \chi_{anal.}\;
  \left(\frac{\coup_c - \coup}{\coup_c} \right)^{-\gamma+1} ,
  \label{correktk}
  \ee
with $\Delta_1 = \omega \nu$, where $\nu$ is the critical exponent of the
correlation length and $\omega$ denotes the correction-to-scaling exponent.
We fitted our data according to the scaling law with
corrections. Since a fit with 6 free parameters is hard to stabilize, we fixed
the
critical exponents to the values $\gamma = 1.3160(25)$,
$\omega = 0.780(25)$ and $\nu = 0.669(2)$ which are obtained from
resummed perturbation series
\cite{guillou80}. Including all the
data points in the  fit we get $K_c = 0.454162(9)$,
$\chi_0 = 1.058(7)$, $\chi_{conf.} = -0.16=(6)$ and $\chi_{anal.} = 0.18(10)$
with $\chi^2/d.o.f. \approx 0.73$.
The $\chi_0$ which is
obtained from the simple scaling fit (\ref{simk}), and that obtained from the
fit
allowing corrections to the scaling, differ by a larger amount than their
statistical errors. This shows that one cannot interpret a small
$\chi^2 / d.o.f.$ as the absence of systematic errors due to an incomplete fit
ansatz.

One can also write the scaling relations in terms of the temperature
$T=\frac{1}{\coup}$. This leads to
  \be
  \widetilde{\chi}(T) = \widetilde{\chi}_0\;\left(\frac{T - T_c}{T_c}
  \right)^{-\gamma} \label{simt}
  \ee
and
  \be
  \widetilde{\chi}(T) =
  \widetilde{\chi}_0        \; \left(\frac{T - T_c}{T_c} \right)^{-\gamma} +
  \widetilde{\chi}_{conf.}\;
    \left(\frac{T - T_c}{T_c} \right)^{-\gamma+\Delta_1}+
  \widetilde{\chi}_{anal.}\;
    \left( \frac{T - T_c}{T_c} \right)^{-\gamma+1}
  \label{correktt}
  \ee
with corrections. We repeated the analysis as done above.
 Taking all 15 data points
into account we get for the simple scaling fit a
 $\chi^2/d.o.f \approx 61.2$. We again subsequently discarded  data points
with small $\coup$. A summary of the results is given in Table\ \ref{chitfit}.
Starting from 5 discarded data points the $\chi^2 / d.o.f.$
 becomes approximately 1. But the results obtained for
$K_c$, $\gamma$ and $\chi_0$ are not consistent with those obtained from the
fit according to eq.\ (\ref{simk}).

Finally, we performed a four-parameter fit to the scaling relation with
corrections and fixed values for the exponents.
Taking all data points into account we obtain
$K_c = 0.454163(9)$,  $\widetilde{\chi}_0 = 1.059(7)$,
$\widetilde{\chi}_{conf.} = -0.17(6)$ and $\widetilde{\chi}_{anal.} = 1.59(10)$
with $\chi^2/d.o.f. \approx 0.71$.
The results for $\coup_c$, $\chi_0$ and $\chi_{conf}$
of the fits according to the ans\"atze (\ref{correktk})
and (\ref{correktt})    are consistent within the error bars. The ambiguity
between the ansatz with $\coup$ as variable and that with $T$ as variable
is covered
by the analytic corrections.

We conclude that the scaling ansatz  (\ref{simk})  fits well if one chooses the
coupling $\coup$ as the variable. But we also learned that a small $\chi^2 /
d.o.f.$ does not exclude systematic errors, due to corrections to the scaling,
that
are larger than the  statistical ones.
Hence it is hard to give  final estimates obtained from the simple scaling
ansatz
that also take systematic errors into account.
 From the ansatz with corrections to the scaling
we obtain,
 assuming $\gamma = 1.3160(25)$ and $ \Delta_1 = 0.52182 $, the results
$K_c =  0.454162(13)$, $\chi_0 =  1.058(22)$ and $\chi_{conf.} = -0.16(11)$,
where the uncertainty of $\gamma$ is taken into account.

We also like to  emphasize that the
Wegner amplitude $\chi_{conf.}$ is negative for the fits that take corrections
into account. This is
in agreement with a field-theoretical renormalization group
calculation of Esser and
Dohm, which predicts the confluent correction-to-scaling amplitude to be
negative for a finite cut-off \cite{esserdohm}.

\subsubsection{Correlation length}
We extracted  $\xi_{exp}$  from the large distance behaviour of
the improved time slice correlation function eq.(\ref{corfunc}).
We therefore considered the effective
correlation length, defined by
  \be
  \xi_{ef\!f}(t)=-\ln\frac{\overline{G}(t-1)}{\overline{G}(t)} \, ,
  \label{effcorr}
  \ee
where for brevity
we have suppressed the contribution due to periodic boundary conditions.
As an example, we show in Fig.\ \ref{effco} the results for
 $ \xi_{ef\!f}(t) $ obtained on a
  $112^3$ lattice at $K=0.452$. A
single state dominates the correlation function and a plateau sets in around
$t=\xi_{exp} /2$ and extends to $t=3\xi_{exp}$, with no visible degradation
due to increasing statistical errors at large $t$.
 As our final estimate for $\xi_{exp}$ we took
 self-consistently $ \xi_{ef\!f}(t) $ at the distance $t = 2 \xi_{exp}$.

In order to
 calculate $\xi_{2^{nd}}$ we used the improved version of eq.(\ref{xi2def}).
The advantage of this definition is that no fit is needed to obtain the
correlation length. The data of $\xi_{exp}$ and $\xi_{2^{nd}}$ are given in
Table\ \ref{corrlength}.
The deviation of $\xi_{2^{nd}}$ from $\xi_{exp}$ is about $1 \%$ for $K=0.40$
and becomes smaller than
 $0.1 \%$ for $K \ge 0.448 $.

Since the difference of $\xi_{exp}$ and $\xi_{2^{nd}}$ is so small, we will
discuss only the results of $\xi_{exp}$ in the following.
 {F}irst we did a three-parameter fit  for
$\xi_{exp}$  following the simple scaling ansatz
  \be
  \xi(K) = \xi_0\;
 \left(\frac{\coup_c - \coup}{\coup_c}\right)^{-\nu}.\label{simxi}
  \ee
The results  are given in Table\ \ref{xiextab}.
Taking all data into account, the fit has a large  $\chi^2/d.o.f$ of about
 9.
Starting with three data points with small $\coup$ being discarded,
  the $\chi^2 / d.o.f.$ is close to 1.
 But still the critical exponent $\nu$ and the critical coupling
systematically tend to smaller values.

If we fit the data to the simple
scaling ansatz (\ref{simxi}), where the coupling is replaced as variable
by the temperature, a similar behaviour is observable. The results
are shown in Table\ \ref{xiexttab}. Here one also has to
discard three data points to obtain a $\chi^2 / d.o.f.$ close to $1$.
But now the estimates of the critical exponent and the critical
coupling start at lower values and tend to larger ones.

This indicates that corrections to the simple scaling ansatz  have to be taken
into account.
Therefore we have fitted all data to the
scaling relation with corrections given by
  \be
  \xi(\coup) = \xi_0\;
  \left(\frac{\coup_c - \coup}{\coup_c} \right)^{-\nu} + \xi_{conf.}\;
  \left(\frac{\coup_c - \coup}{\coup_c} \right)
  ^{-\nu+\Delta_1}+ \xi_{anal.}\; \left(\frac{\coup_c - \coup}{\coup_c}
 \right)^{-\nu+1},
  \ee
whereas, in the case of the magnetic susceptibility, $\Delta_1 = \omega\nu$.
The four-parameter fit to all data points
with the critical exponents fixed to the resummed perturbation series
 estimates leads
to $K_c = 0.454167(10)$, $\xi_0 = 0.498(2)$, $\xi_{conf.} = -0.10(2)$ and
$\xi_{anal.} = -0.07(4)$ with $\chi^2/d.o.f. \approx 0.63$.

We also made a four-parameter fit to the scaling relation with corrections
where the coupling is replaced by the temperature. This leads
to $K_c = 0.454165(10)$,   $\widetilde{\xi}_0 = 0.498(2)$,
$\widetilde{\xi}_{conf.} = -0.09(2)$, $\widetilde{\xi}_{anal.} = 0.24(3)$
with $\chi^2/d.o.f. \approx 0.63$.

Taking the uncertainty of $\nu$ into account, we leave at $K_c =
0.454166(15)$, $\xi_0 = 0.498(8)$ and $\xi_{conf.} = -0.10(6)$.

In summary, we conclude that systematic deviations from the simple scaling
ansatz (\ref{simxi}) due to corrections to scaling are important for the
analysis
of the correlation length data in the coupling range that is accessible to
Monte Carlo simulations. Thus it is hard to obtain
accurate estimates for the critical exponents and the critical coupling from
such an approach.

\section{Numerical results at criticality}
On lattices of the size $L =  4,8,16,32 $ and 64 we performed simulations
at $K_0=0.45417$ which is the estimate for the critical coupling
obtained in the previous section.  As in the high temperature simulation
the statistics are given in terms of $N$ measurements taken every $N_0$
update steps.  We have chosen $N_0$ such  that on the average
the lattice is updated approximately twice for one  measurement.
 The results of the runs are
summarized in Table\ \ref{crresult}.
\subsection{Phenomenological Renormalization Group}
First we determined the critical coupling $K_c$ and the
critical exponent $\nu$ employing Binder's phenomenological renormalization
group method \cite{binder81b}.
 In addition to the fourth-order cumulant   defined on the whole
 lattice we also studied cumulants defined on subblocks of the lattice.
Therefore let us first introduce blockspins
  \be
  S_B = L_B^{1/2(D-2)}\;\frac{1}{{L_B}^D}\;\sum_{i\in B} \vec{s}_i\; ,
  \ee
where $L_B$ is the size of the block and $1/2(D-2)$ is the canonical dimension
of the field.
 In particular we studied the fourth-order cumulant
  \be
  U_{L_B} = 1 - \frac{<(S_B^2)^2>}{3<S_B^2>^2}
  \ee
for $L_B=L, L/2$ and a nearest neighbour interaction on subblocks
  \be
  N\!N = \frac{<S_{B_1}S_{B_2}>}{<S_B^2>}
  \ee
for $L_B = L/2$.

For the extrapolation to couplings $K$ other than the simulation coupling
$K_0$,
we used the reweighting formula 
  \begin{equation}
   \langle A \rangle (K) = \frac{\sum_i A_i \exp((-K+K_0) H_i)}
                     {\sum_i  \exp((-K+K_0) H_i)} ,
  \label{reweight}
  \end{equation}
where $i$ labels the configurations generated according to the Boltzmann-weight
at $K_0$. We computed the statistical errors from Jackknife binning on the
final result of the extrapolated cumulants. The extrapolation only gives good
results within a small neighbourhood of the simulation coupling $K_0$. This
range shrinks with increasing volume of the lattice.

For sufficiently large $L_B$ the cumulants have a
non-trival fixed point  at the critical coupling. When one considers the
cumulants as a function  of the coupling, the crossings of the curves for
different $L$
provide  an estimate for the critical coupling $K_c$.  As an example we show
in Fig.\ \ref{histou1} the fourth-order cumulant in a
neighbourhood of $K_c$. The figure shows that the crossings of the cumulant
are well covered by the extrapolation (\ref{reweight}). The error bars of
$U_L$ with $L = 64$ blow up for $|K-K_0| > 0.001 $ while
$|K_{cross} - K_0| = 0.00003$ for $L=32$ and $L=64$.
The results for the crossings are summarized in Table\ \ref{coupinter}.
The given errors are taken from the size of the crossings of the error bars.
The convergence of the crossing coupling $K_{cross}$ towards $K_c$ should
follow
  \be
  K_{cross}(L)  =  K_c \;( 1 + const.  L^{ -(\omega+\frac{1}{\nu})}),
  \label{kcross}
  \ee
where $\omega$ is the  correction to scaling exponent \cite{binder81b}.
Our data for the crossings of the cumulants did not allow
us to perform a two-parameter fit, keeping the exponents fixed,
following the above formula.  Within the statistical errors the results of
the crossings of the fourth-order cumulants
on $L=8$ and $L=16$ up to $L=32$ and $L=64$ are consistent.
The convergence of the crossings of $N\!N$ towards $K_c$ seems to be slower
than that of the fourth-order cumulants, but
it is interesting to note that the $K_{cross}$
for the fourth-order cumulant and
$N\!N$  come from different  sides with increasing  $L$.
This is shown in Fig.\ \ref{critcoups}, where the estimates of $K_{cross}$
versus the lattice size $L$ are plotted.
Our final estimate for the critical coupling
is $K_c = 0.45420(2)$ obtained from the  $L=32$ and $L=64$ crossing
of the fourth-order  cumulant on the full lattice.
Taking into account the fast convergence of the crossings towards $K_c$, that
is predicted by (\ref{kcross}),  we conclude that the systematic error of our
estimate for $K_c$  is smaller than the given statistical error.

At the critical coupling
$K_c$ the cumulants converge with increasing lattice size $L$
 to a universal fixed point. The convergence rate is given by \cite{binder81b}
  \be
  U_{L}(K_c)  =  U_{\infty} \;( 1 + const.  L^{ -\omega}) \, .
  \label{ufix}
  \ee
 The results for the  cumulants at $K=0.45420$ , which is our estimate of
critical coupling, are given in Table\ \ref{renobs}. The data did not allow
us to perform a two parameter fit with $\omega$ being fixed.
Hence we take the value
$U_{L}(K_c) = 0.589(2)$ form $L=64$
 as our  final estimate for the fixed point of the fourth-order cumulant on the
full
lattice,
 where we now have taken into account the
uncertainty of the estimated critical coupling.

We extracted the critical exponent  $\nu$ of the correlation length from the
$L$ dependence
of the slope of the fourth-order cumulant at criticality \cite{binder81b}.
According to Binder, the scaling relation for the slope of the fourth-order
cumulant is given by
  \be
  \left .\frac{\partial U(L,\coup)}{\partial\coup} \right |_{\coup_c}  \propto
  L^{1/\nu}.
  \label{slope}
  \ee
We evaluated the slopes of the observables $A$ entering  the cumulant $U$
according to
  \be
  \frac{\partial \langle A \rangle }{\partial \coup} = \langle A H \rangle -
  \langle A \rangle \langle H \rangle ,
  \label{steig}
  \ee
where $A$ is an observable and $H$ is the energy.
The statistical errors are calculated from  a Jackknife analysis for the value
of the slope. First we estimated the exponent $\nu$ from different lattices via
  \be
  \nu = \frac{\ln \left( L_2 \right)- \ln\left(L_1\right)}{\displaystyle
  \ln \left(
  \left.\frac{\partial A(L_2,\coup)}{\partial\coup} \right |_{\coup_c}\right)-
  \ln \left(\left.\frac{\partial A(L_1,\coup)}{\partial\coup} \right
  |_{\coup_c}\right)}.
  \label{nulat}
  \ee
The results are given in Table\ \ref{nutab}.
The estimates for $\nu$ stemming from  $U_L$ and $U_{L/2}$
are stable with increasing $L$ and consistent with each other for  $L \ge 16$.
Therefore we performed a fit according to  eq.(\ref{slope}) with $U_L$
from  lattices of the size $L=16$ up to $L=64$.
We consider the result $\nu=0.662(7)$ as our final estimate for the critical
exponent of the correlation length.

\subsection{Magnetic Susceptibility}
In order to estimate the ratio $\gamma/\nu$ of the critical exponents we
studied
the scaling behaviour of the magnetic susceptibility
defined on the  full lattice and on subblocks.
The dependence of the susceptibility on the lattice size
is given by
 \be
 {\chi} \propto L^{\gamma / \nu}
 \label{h}
 \ee
at the critical coupling.
We have estimated $\gamma/\nu$ from pairs of lattices with size $L_1$,
$L_2$. The ratio is then given by
  \be
  \frac{\gamma}{\nu} = \frac{\ln(\chi(L_1,\coup_c))-\ln(\chi(L_2,\coup_c))}
  {\ln(L_1)-\ln(L_2)} .  \label{phen}
  \ee
The second column of Table\ \ref{gamnu} shows the
estimates of the ratio obtained from the susceptibility defined on  the full
lattice, while the third column shows the estimates
obtained from the blockspin-susceptibility with subblocks of the size $L/2$.
The estimates for $\gamma / \nu$ obtained
 from the subblocks monotonically increase
with increasing lattice size $L$, while those obtained from the full lattice
decrease. The results obtained from the full lattice for $L \ge 16$ and
the result from the subblocks of the largest lattices are consistent within
error bars.  Hence we take
$\gamma / \nu=1.976(6)$ as our final result, where
statistical  as well as systematic errors should be covered.
 Using the scaling relation
$\eta = 2 - \frac{\gamma}{\nu}$
we obtain for the anomalous dimension $\eta=0.024(6)$.

\subsection{Hyperscaling and specific heat}
In ref. \cite{binder81b}
a dangerous irrelevant scaling field $u$ is proposed  as explanation for a
possible violation of hyperscaling. Dangerous means that the scaling function
of the correlation length  vanishes with some power $q$ of the vanishing
irrelevant scaling field. Hence the correlation length should scale as
  \be
  \xi \propto L^{1+q  y_u}
  \ee
at the critical point.
 Remember that $y_u$ is negative for an irrelevant scaling
field.
 At $\coup_c=0.45420$, which we obtained from the analysis of the fourth-order
cumulant, we have fitted $\xi_{2^{nd}}$ to this relation.
The reweighted estimates of $\xi_{2^{nd}}$ are shown in Table\ \ref{critres}.
Taking lattices
of size $L=16$ up to $L=64$ into account we estimate $qy_u = 0.007(2)$ with
$\chi^2/d.o.f. \approx 0.3$ and only statistical errors considered. This
indicates that there is no or only very small hyperscaling violation due to a
dangerous irrelevant field.

At criticality the specific heat  should scale as
  \be
  C_h(L) =  C_{reg} + const.\;  L^{\frac{\alpha}{\nu}},
  \ee
where $C_{reg}$ denotes the regular part of the specific heat and $\alpha$ is
the critical exponent of the specific heat. Using the critical exponent
$\nu=0.662(7)$ obtained from the analysis above the hyperscaling relation
$\alpha = 2 - D \nu$ gives $\alpha = 0.014(21)$. We also tried to estimate
$\alpha$ via a three-parameter fit, following the finite-size scaling
relation. However, we are not able to give a stable estimate for $\alpha$.

\subsection{Helicity modulus}
The 3$D$ $XY$ model is assumed to share
the same universality class as an interacting
Bose fluid, and the helicity modulus $\heli$
should be proportional to the superfluid density $\densi$ of the Bose
fluid \cite{fisher73a}. Near the
critical coupling the superfluid density, resp. the helicity modulus should
scale as
  \be
  \densi \propto \heli \propto \left | \coup-\coup_c\right |^v,
  \ee
with $v$ the critical exponent of the superfluid density. Assuming
hyperscaling the Josephson relation reads $v = (D - 2) \nu $ \cite{fisher73a}.
Hence the product
  \be
  \heli \cdot L =const
  \ee
should stay constant at the critical point in 3$D$.
To check this prediction we have measured the helicity modulus $\heli$ on
lattices of size $L=4$ to $L=32$. The estimator of $\heli L$ becomes noisy
with increasing lattice size.  We tried to overcome this problem by measuring
more often, which did not remove the problem  completely. Hence
we skipped the measurement of $\heli$ for $L=64$. The results, shown in Table\
\ref{critres}, indicate that the above relation holds.

\subsection{Performance of the Algorithm}

The efficiency of a stochastic algorithm is characterized by the
autocorrelation time
   \begin{equation}
   \tau ={1 \over 2} \sum_{t = - \infty}^{\infty} \rho (t) \, ,
   \label{f}
   \end{equation}
where  the normalized autocorrelation function $\rho (t)$ of an observable
$A$ is given by
   \begin{equation}
   \rho (t) = {{ \langle A_i \cdot A_{i+t} \rangle - {\langle A \rangle}^2}
   \over{\langle A^2 \rangle - {\langle A \rangle}^2 }} \,  .
   \label{e}
   \end{equation}
We  calculated the integrated autocorrelation times
$\tau$ with a self-consistent truncation window of width 6$\tau$
for the energy density $E$ and the magnetic susceptibility $\chi$ for lattices
with $L = 4$ up to $L=64$ at the  coupling $\coup=0.45417$.
In Fig.\ \ref{zfit} we show a log-log plot of the integrated autocorrelation
times $\tau$ of the energy density $E$ and the magnetic susceptibility $\chi$
versus the  lattice size $L$ given in units of the
 average number of clusters that
is needed to cover the volume of the lattice.
Our estimates for the critical dynamical exponents are $z_{E} = 0.21(1)$ and
$z_{\chi}=0.07(1)$ taking only statistical errors into account. These results
are consistent with those of Janke \cite{janke90a}.

Finally let us briefly comment on the CPU time:
160 single cluster updates of the $64^3$ lattice at the  coupling
$K = 0.45417$ plus one measurement of
the observables took on average 26 sec CPU time on a IBM RISC 6000-550
workstation.
All our MC simulations of the 3D XY model together
took about two months of CPU-time.

\section{Comparison of our results with previous studies}
In Table\ \ref{res} we display estimates of critical properties of the $3D$
$XY$-model obtained  by various methods. Our estimates of $K_c$ from the
scaling fit to the high temperature data and from the penomenological RG
approach are consistent within 2 standard deviations. But only for the result
from the phenomenological RG approach are the systematical errors fully under
control. Our error of $K_c$ is about 4 times smaller than that of previous
MC studies \cite{we90a,janke90a}, and also
about 4 times smaller than that obtained recently \cite{Adler}
from the analysis of a $14th$ order high temperature series
expansion \cite{luescher}. Recently Butera et al. \cite{butera} extended
the high temperature series expansion for the sc lattice to the order $17$.
Their value for the critical coupling is by three times their error estimate
smaller than our value.

The error of $\gamma$ obtained from a fit to the simple scaling ansatz is
about 5 times smaller than those of previous MC studies\cite{we90a,janke90a};
however, the systematical errors are not under control. The value of $\gamma$
is, within two  standard deviations, consistent with the estimate of
Ref.\cite{janke90a}.
Our estimate of $\gamma$ is consistent with the value obtained from the high
temperature series expansion \cite{ferrer73a,Adler,butera}
 and, within two standard deviations,
consistent with the value of the $\epsilon$-expansion \cite{guillou85a} while
the very accurate
estimate from the resummed perturbation series \cite{guillou80}  is
smaller than our estimate by three times
the quoted error.

Our estimate for $\nu$ is consistent within error bars with all other estimates
we quote in Table\ \ref{res}. Our quoted error bars are 3.5 times larger than
that of ref. \cite{janke90a}. Janke used finite differences  to determine
the slope of the cumulant \cite{privat},
 while we used fluctuations at a single temperature
(\ref{steig}). Furthermore the smallest lattice size $L=16$ included in our fit
 is chosen to be rather conservative.
The most accurate number for $\nu$ stems from the measurement of the
superfluid fraction of $^4$He \cite{ahlers}.

In this work we give for the first time an accurate direct
MC estimate for the exponent
$\eta$, the anomalous dimension of the field. The uncertainty of the
estimate is comparable  with those  obtained with field theoretical
methods. Our value of the exponent $\eta$ is consistent with the estimates
from the high temperature series expansion \cite{ferrer73a,Adler} and
with that of the resummed perturbation series \cite{guillou80},
 but is
smaller than the $\epsilon$-expansion \cite{guillou85a} result
by more than twice our error estimate.

Our result for the critical fourth-order cumulant, is consistent
with previous MC results \cite{we90a,janke90a}. But the value obtained
from $\epsilon$-expansion \cite{brezin85a}
is off by about 20 times our error estimate that
also takes into account systematic errors.
Furthermore we provide estimates for the critical fourth order cumulant
on subblocks and a nearest neighbour blockspin product $N\!N$. These numbers
might be useful in testing other models sharing the $XY$ universality class.

\section{Conclusions}
The application
 of the  single cluster algorithm \cite{wolff89a},
 which is almost free of critical
slowing down for the $3D$ $XY$ model, and the extensive use of
modern RISC workstations allowed us to increase the statistics as well
as the studied lattices sizes considerably compared with
 previous MC simulations
\cite{we90a,janke90a}.
In the high temperature phase of the model we measured correlation length
up to  17.58 with an accuracy of about 0.1$\%$. But the analysis of our data
for the correlation length and the magnetic susceptibility showed that it
is hard to control systematic errors due to confluent and analytic
 corrections.
It seems to be much easier to fight the  systematic errors in
the phenomenological RG approach. Analytic corrections are absent at the
critical point and corrections to the scaling are less harmful, since the
relevant length scale at criticality is the lattice size,  which can be chosen
much larger than the correlation length  in the thermodynamic limit of the
high temperature phase.
{}From the crossings of the fourth-order cumulant we
obtain $K_c = 0.45420(2) $, which reduces  the error by a factor of
about 4 compared with previous MC studies \cite{we90a,janke90a}.
Further improvements of the accuracy of the estimates of the critical
coupling and critical exponents seem to be reachable by just increasing
the statistics, while keeping the present lattice sizes.
The accurate values obtained for critical cumulants
could be very useful for testing
whether other models  share the $XY$ universality class. Here of
course a proper block-spin definition is essential.

\acknowledgments
The numerical simulations were performed on an IBM RISC 6000 cluster of the
Regionales Hochschulrechenzentrum Kaiserslautern (RHRK).
The work was supported in part by Deutsche Forschungsgemeinschaft (DFG)
under grant Me 567/5-3. It is a pleasure to thank S. Meyer for discussions.
We would like to thank W. Janke for communicating to us some unpublished
details of his study.


%

\newpage

%
\begin{figure}
\caption{
The effective correlation length $\xi_{e\!f\!f}(t)$ as a function of
separation  for coupling $\coup=0.452$ on a lattice of size $L = 112 $.
\label{effco}}
\end{figure}

%
\begin{figure}
\caption{
Reweighting plot of the Binder cumulant $U_L$ of the full lattice
from the simulation at $\coup=0.45417$. The dashed lines give the statistical
errors obtained by a binning analysis.
\label{histou1}}
\end{figure}

%
\begin{figure}
\caption{
Plot of the convergence of the critical coupling obtained by the cumulant
crossing method. Because of the small statistical errors one  is able to see
systematic convergence of the critical coupling.
\label{critcoups}}
\end{figure}

%
\begin{figure}
\caption{
Intergated autocorrelation times $\tau$ of the energy
density $E$ and the magnetic susceptibility $\chi$ versus the  lattice size
$L$.  The dynamical critical exponent is given by the slopes of the
fits.
\label{zfit}}
\end{figure}
%

\newpage
\pagestyle{revtex}

%
\begin{table}
\caption{
Results of the energy density $E$, the specific heat at constant external
field $C_h$, the impoved susceptibility $\chi_{imp}$ and the standard
susceptibility $\chi$ obtained from the simulations in the high temperature
phase. The parameters of the runs are given in terms of ~the simulation
coupling
$\coup$, the linear size of the system $L$, and the statistics, with  $N$
the number of measurements taken every $N_0$ update steps.
}
\label{htresult}
\begin{tabular}{crccllll}
\multicolumn{1}{c}{ $\coup$} &
\multicolumn{1}{c}{ $L$}    &
 $N$ & $N_0$&
\multicolumn{1}{c}{ $E$}    &
\multicolumn{1}{c}{ $C_h$}  &
\multicolumn{1}{c}{ $\chi_{imp}$}  &
\multicolumn{1}{c}{ $\chi$}   \\ \hline
0.400 & 24 & 20k & 0.9k& 0.24697(6) &  3.17(3) &  \phantom{11}16.848(14) &
\phantom{11}16.70(12)  \\

0.410 & 32 & 10k & 2k & 0.25779(5) &  3.39(6) &  \phantom{11}22.108(19)  &
\phantom{11}22.24(32)   \\

0.420 & 32 & 10k & 1.6k& 0.26970(5) &  3.65(6) &  \phantom{11}30.994(29) &
\phantom{11}31.18(32) \\

0.425 & 32 & 10k & 1.2k& 0.27605(6) &  3.96(7) &  \phantom{11}38.21(5)  &
\phantom{11}38.47(38) \\

0.430 & 32 & 20k & 0.8k& 0.28286(4) &  4.22(4) &  \phantom{11}49.12(6)  &
\phantom{11}48.87(36) \\

0.435 & 32 & 20k & 0.6k& 0.29023(7) &  4.57(8) &  \phantom{11}66.57(15)  &
\phantom{11}66.6(7) \\

0.437 & 32 & 20k & 0.4k& 0.29329(5) &  4.72(6) &  \phantom{11}77.11(16)  &
\phantom{11}76.9(5) \\

0.440 & 32 & 20k & 0.4k& 0.29818(5) &  5.11(6) &  \phantom{11}99.50(20)  &
\phantom{11}99.2(7) \\

0.443 & 48 & 21k & 1k  & 0.30336(3) &  5.39(7) & \phantom{1}136.41(20)   &
\phantom{1}135.8(9) \\

0.445 & 48 & 40k & 0.5k& 0.30705(3) &  5.84(6) & \phantom{1}177.04(30)   &
\phantom{1}176.2(9) \\

0.448 & 64 & 25k &  1k & 0.31303(2) &  6.43(8) & \phantom{1}299.14(53)   &
\phantom{1}301.0(2.0) \\

0.449 & 64 & 25k &  1k & 0.315214(18)& 6.61(7) & \phantom{1}377.75(64)   &
\phantom{1}378.9(2.5) \\

0.450 & 96 & 12k &  3k & 0.317497(15)& 7.02(11)& \phantom{1}503.85(85)    &
\phantom{1}506.5(4.6) \\

0.451 & 96 & 12k &  2k & 0.319854(13)& 7.33(10)& \phantom{1}722.1(1.3)    &
\phantom{1}724.6(6.5) \\

0.452 &112 & 12k &  2k & 0.322446(12)& 8.10(14)&   1193.0(3.0)            &
1201.0(10.0) \\
\end{tabular}
\end{table}

%
\begin{table}
\caption{
Results for  the correlation lengths $\xi_{exp}$ and $\xi_{2^{nd}}$ obtained
from
the simulations in the high temperature phase. $\xi_{2^{nd}}/\xi_{exp}$ gives
the ratio between the  values of the two correlation lengths.
}
\label{corrlength}
\begin{tabular}{crrrl}
$\coup$ & \multicolumn{1}{c}{ $L$} &
\multicolumn{1}{c}{$\xi_{2^{nd}}$} &
\multicolumn{1}{c}{$\xi_{exp}$} & $\xi_{2^{nd}}/\xi_{exp}$ \\ \hline

0.400 & 24 &  1.876(1) & 1.898(2) & 0.9884\\

0.410 & 32 &  2.182(1) & 2.202(2) & 0.9909\\

0.420 & 32 &  2.624(2) & 2.639(2) & 0.9943\\

0.425 & 32 &  2.938(2) & 2.953(2) & 0.9949\\

0.430 & 32 &  3.361(2) & 3.375(3) & 0.9959\\

0.435 & 32 &  3.947(5) & 3.959(5) & 0.9969\\

0.437 & 32 &  4.262(5) & 4.273(6) & 0.9974\\

0.440 & 32 &  4.875(5) & 4.885(6) & 0.9979\\

0.443 & 48 &  5.746(5) & 5.756(6) & 0.9983\\

0.445 & 48 &  6.582(6) & 6.593(7) & 0.9983\\

0.448 & 64 &  8.638(9) & 8.645(10) & 0.9991\\

0.449 & 64 &  9.738(9) & 9.747(10) & 0.9991\\

0.450 & 96 & 11.288(10) &11.295(10) & 0.9994\\

0.451 & 96 & 13.587(16) &13.594(16) & 0.9995\\

0.452 &112 & 17.570(19) &17.580(20) & 0.9994\\
\end{tabular}
\end{table}

%
\begin{table}
\caption{
Results of the energy density $E$, the specific heat $C_h$, the  helicity
modulus $\heli$, the improved susceptibility $\chi_{imp}$, the correlation
lengths $\xi_{exp}$ and $\xi_{2^{nd}}$ and the ratio $L/\xi_{2^{nd}}$ obtained
from simulations at $\coup=0.435$ with linear system size $L$. The statistics
is given in terms of $N$ measurements taken every $N_0$ update steps.
}
\label{fsizetab}
\begin{tabular}{cccccccllc}
$L$ &
$N$ & $N_0$ &
$E$   &
$C_h$ &
$\heli$ &
$\chi_{imp}$ &
\multicolumn{1}{c}{$\xi_{2^{nd}}$} &
\multicolumn{1}{c}{$\xi_{exp}$} &
$L/\xi_{2^{nd}}$ \\
\hline
4   &  25k  & 0.1k  &  0.3621(26) & 6.43(13) &   0.2180(44) &  15.83(21)  &
    2.015(19) & 2.08(2)   &    1.98   \\
8   &  20k  & 0.2k  &  0.3054(26) & 6.56(13) &   0.0608(16) &  43.00(07)  &
    3.228(3)  & 3.272(4)  &    2.48    \\
16  &  20k  & 0.2k  &  0.29114(9) & 4.92(6)  &   0.0056(17) &  64.31(16)  &
    3.886(6)  & 3.901(6)  &    4.12   \\
20  &  20k  & 0.3k  &  0.29049(7) & 4.71(5)  &   0.0031(16) &  66.21(14)  &
    3.938(4)  & 3.949(5)  &    5.08   \\
24  &  20k  & 0.3k  &  0.29024(7) & 4.65(5)  &$-$0.0001(17) &  66.55(15)  &
    3.946(6)  & 3.959(6)  &    6.08   \\
32  &  10k  & 0.6k  &  0.29013(7) & 4.52(9)  &   0.0009(25) &  66.64(15)  &
   3.949(5)  & 3.960(5)   &    8.10   \\
\end{tabular}
\end{table}

%
\begin{table}
\caption{
Estimates for the critical coupling
$\coup_c$, the static critical exponent $\gamma$
and the amplitude $\chi_0$ obtained from a fit of the improved susceptibility
$\chi_{imp}$ to eq.(\protect\ref{simk}).
$\chi^2/d.o.f$ gives the quality of
the fit. \# denotes the number of discarded data points at small couplings.
}
\label{chikfit}
\begin{tabular}{rllll}
\#&     $K_c$      &    $\gamma$     &    $\chi_0$   &$\chi^2/d.o.f.$\\ \hline
0  & 0.454170(7) & 1.3241(10)  &  1.0090(24) & 0.65 \\
1  & 0.454168(8) & 1.3238(12)  &  1.0099(30) & 0.69 \\
2  & 0.454175(9) & 1.3252(14)  &  1.0057(40) & 0.49 \\
3  & 0.454173(10)& 1.3248(18)  &  1.0069(53) & 0.53 \\
4  & 0.454170(11)& 1.3239(22)  &  1.0101(66) & 0.52 \\
5  & 0.454179(14)& 1.3264(31)  &  1.0016(100)& 0.41 \\
6  & 0.454176(15)& 1.3256(35)  &  1.0043(117)& 0.44 \\
7  & 0.454174(17)& 1.3251(43)  &  1.0060(145)& 0.52 \\
8  & 0.454174(19)& 1.3250(52)  &  1.0063(180)& 0.65 \\
9  & 0.454180(24)& 1.3272(74)  &  0.9979(267)& 0.81 \\
10 & 0.454197(42)& 1.3337(156) &  0.9736(576)& 1.11 \\
11 & 0.454208(58)& 1.3384(240) &  0.9557(885)& 2.14 \\
\end{tabular}
\end{table}

%
\begin{table}
\caption{
Estimates for the critical coupling $\coup_c$,
the static critical exponent $\gamma$
and the amplitude $\chi_0$ obtained from a fit of the improved susceptibility
to eq.(\protect\ref{simt}).
 $\chi^2/d.o.f$ gives the quality of the fit. \# denotes
the number of discarded data points at small couplings.
}
\label{chitfit}
\begin{tabular}{rllll}
\#&     $K_c$   &  $\gamma$  & $\widehat{\chi}_0$ & $\chi^2/d.o.f.$\\ \hline
0 & 0.453871(6) & 1.2351(8)  & 1.3972(27) & 77.1 \\
1 & 0.453932(7) & 1.2471(10) & 1.3500(34) & 40.0 \\
2 & 0.453995(8) & 1.2604(13) & 1.2970(45) & 12.8 \\
3 & 0.454028(9) & 1.2683(16) & 1.2648(59) & 6.98 \\
4 & 0.454040(10)& 1.2733(20) & 1.2440(74) & 5.27 \\
5 & 0.454091(13)& 1.2849(29) & 1.1959(111)& 1.39 \\
6 & 0.454098(14)& 1.2871(33) & 1.1870(128)& 1.31 \\
7 & 0.454110(16)& 1.2907(40) & 1.1720(158)& 1.08 \\
8 & 0.454120(18)& 1.2938(49) & 1.1587(196)& 1.02 \\
9 & 0.454135(23)& 1.2991(70) & 1.1359(289)& 0.99 \\
10& 0.454166(42)& 1.3112(150)& 1.0845(618)& 1.06 \\
11& 0.454182(59)& 1.3177(231)& 1.0570(945)& 1.98 \\
\end{tabular}
\end{table}

\begin{table}
\caption{%
Estimates for the critical coupling
$\coup_c$, the static critical exponent $\nu$
and the amplitude $\xi_0$ obtained from a fit of the exponential correlation
length $\xi_{exp}$ to eq.(\protect\ref{simxi}).
$\chi^2/d.o.f$ gives the quality of
the fit. \# denotes the number of discarded data points at small couplings.
}
\label{xiextab}
%
\begin{tabular}{rllll}
\#&     $K_c$      &    $\nu$     &    $\xi_0$   &$\chi^2/d.o.f.$\\ \hline
 0  &  0.454325(9) &  0.7029(7) & 0.4294(8) &  9.2 \\
 1  &  0.454301(9) &  0.7003(8) & 0.4327(9) &  4.9  \\
 2  &  0.454286(10)&  0.6985(9) & 0.4351(11)&  3.6 \\
 3  &  0.454269(11)&  0.6964(10)& 0.4381(13)&  2.0 \\
 4  &  0.454247(12)&  0.6933(13)& 0.4426(18)&  0.66 \\
 5  &  0.454243(14)&  0.6927(18)& 0.4436(25)&  0.71 \\
 6  &  0.454235(16)&  0.6914(21)& 0.4457(31)&  0.59 \\
 7  &  0.454223(18)&  0.6895(24)& 0.4487(37)&  0.31 \\
 8  &  0.454216(20)&  0.6882(30)& 0.4509(48)&  0.25 \\
 9  &  0.454210(24)&  0.6870(39)& 0.4529(64)&  0.26 \\
10  &  0.454208(46)&  0.6866(90)& 0.4537(157)& 0.39 \\
11  &  0.454218(66)&  0.6890(140)& 0.4493(246) & 0.74 \\
\end{tabular}
\end{table}

\begin{table}
\caption{
Estimates for the critical coupling $\coup_c$,
the static critical exponent $\nu$
and the amplitude $\widetilde{\xi}_0$ obtained from a fit of the exponential
correlation length $\xi_{exp}$ to eq.(\protect\ref{simxi}), where the coupling
is replaced by the inverse temperature. $\chi^2/d.o.f$ gives the quality of the
fit. \# denotes the number of discarded data points at small couplings.
}
\label{xiexttab}
%
\begin{tabular}{rllll}
\#  &    $K_c$     &   $\nu$   &$\widetilde{\xi}_0$&   $\chi^2/d.o.f.$ \\
\hline
0   & 0.454079(8)  & 0.6612(6) &  0.5024(8) & 8.5 \\
1   & 0.454098(8)  & 0.6632(7) &  0.4994(9) & 5.3 \\
2   & 0.454118(9)  & 0.6656(8) &  0.4957(11)& 1.7 \\
3   & 0.454125(10) & 0.6665(9) &  0.4944(13)& 1.5 \\
4   & 0.454134(11) & 0.6677(12)&  0.4923(19)& 1.4 \\
5   & 0.454157(14) & 0.6711(16)&  0.4866(26)& 0.18\\
6   & 0.454160(15) & 0.6717(19)&  0.4857(31)& 0.16\\
7   & 0.454161(17) & 0.6717(23)&  0.4856(38)& 0.19\\
8   & 0.454164(20) & 0.6723(29)&  0.4846(49)& 0.22\\
9   & 0.454165(24) & 0.6724(38)&  0.4844(66)& 0.29\\
10  & 0.454177(45) & 0.6750(90)&  0.4796(160)& 0.38\\
11  & 0.454192(65) & 0.6783(136)& 0.4734(250)& 0.65\\
\end{tabular}
\end{table}

%
\begin{table}
\caption{
Results of the energy density $E$, the specific heat $C_h$, the susceptibility
$\chi$ and the second moment correlation length $\xi_{2^{nd}}$ obtained from
simulations at the fixed coupling $\coup = 0.45417$ near the final estimate of
the critical coupling. $\tau$ denotes the integrated autocorrelation time of
the specified observable, given in units of the
average number of clusters
needed to cover the volume of the lattice. The statistics is given in terms
of $N$ measurements taken every $N_0$ update steps.
}
\label{crresult}
\begin{tabular}{lccllllll}
$L$    & $N$ & $N_0$ & $E$        & $\tau_{E}   $
      &  $C_{h}    $  & $\chi    $  &$\tau_{\chi}$ & $\xi_{2^{nd}}$  \\ \hline
  4  & 100k &  10  & 0.40440(44)  &  2.0(1)  & 6.561(27)  &  19.095(34)
  & 1.84(5) & \phantom{1}2.3104(37)\\
  8  &  95k &  20  & 0.35585(20)  &  2.4(1)  & 8.890(39)  &  77.80(15)
  & 1.97(5) &  \phantom{1}4.6852(65)\\
 16  & 100k &  40  & 0.338945(7)  &  2.6(1)  & 10.757(66) &  309.95(60)
  & 1.96(7) &  \phantom{1}9.447(15)\\
 32  &  83k &  80  & 0.332815(3)  &  3.1(1)  & 12.520(73) & 1216.0(2.7)
  & 2.11(5) & 18.922(38)\\
 64  &  72k &  160 & 0.330628(2)  &  3.7(1)  & 14.35(11)  & 4732(12)
  & 2.32(7) & 37.793(77)\\
\end{tabular}
\end{table}

%
\begin{table}
\caption{
Estimates for $K_c(L)$ obtained via Binder's cumulant crossing technique of the
reweighted fourth-order cumulants $U_L$ and $U_{L/2}$ and nearest neighbour
observable $N\!N$. $L_1 - L_2$ gives the pair of linear lattice sizes which
determine the intersection point.
}
\label{coupinter}
\begin{tabular}{cccc}
\multicolumn{4}{c}{$\coup_c(L)$}\\ \hline
 $L_1 - L_2$ & $U_L$ & $U_{L/2}$& $N\!N$\\ \hline
$  4 - 8 $  & 0.4565(4)  & 0.4617(3)  &  0.4378(4)   \\
$  8 - 16$  & 0.4544(2)  & 0.45457(8) &  0.4517(1)   \\
$ 16 - 32$ & 0.45423(5) & 0.45424(4) &  0.45393(4)  \\
$ 32 - 64$ & 0.45420(2) & 0.45421(2) &  0.45415(2)  \\
\end{tabular}
\end{table}

%
\begin{table}
\caption{
Results for the fourth-order cumulants $U_{L}$,  $U_{L/2}$ and the
nearest neighbour observable $N\!N$ at $K = 0.45420$ obtained with
the reweighting technique from the simulations at $K = 0.45417$. The
errors are obtained by a Jackknife-blocking procedure.
}
\label{renobs}
\begin{tabular}{rccc}
  $L$  &    $U_{L}$    &  $U_{L/2}$   &    $N\!N$     \\ \hline
   4   &  0.59640(42)  &  0.56860(25) &  0.70557(61)  \\
   8   &  0.59134(42)  &  0.55270(31) &  0.77439(45)  \\
  16   &  0.58966(43)  &  0.55040(32) &  0.79925(44)  \\
  32   &  0.58907(50)  &  0.54974(37) &  0.80640(48)  \\
  64   &  0.58909(44)  &  0.54925(33) &  0.80963(49)
\end{tabular}
\end{table}

%
\begin{table}
\caption{
Estimates for the static critical exponent $\nu$ obtained using
eq.(\protect\ref{nulat}), where $A$ is replaced by the fourth-order cumulants
$U_L$ and $U_{L/2}$ and the nearest neighbour observable $N\!N$ with the
critical coupling is set to $\coup_c=0.45420$, the final estimate of
the critical coupling.
}
\label{nutab}
\begin{tabular}{cccc}
 lattice& \multicolumn{2}{c}{$\nu$}   \\ \hline
$L_1 - L_2$&$U_L$ &$ U_{L/2}$& $N\!N$ \\ \hline
$  4 - 8 $ & 0.6496(93)  & 0.5807(50)   & 0.8443(86)  \\
$  8 - 16$ & 0.6799(111) & 0.6576(74)   & 0.7519(76)  \\
$ 16 - 32$ & 0.6649(126) & 0.6694(84)   & 0.6977(68)  \\
$ 32 - 64$ & 0.6584(154) & 0.6565(103)  & 0.6779(83)  \\
\end{tabular}
\end{table}

%
\begin{table}
\caption{
Estimates for the ratio of the static critical exponents  $\gamma/\nu$
obtained using eq.(\protect\ref{phen}). The first column gives the results of
the
ratio obtained from the susceptibility of the full lattice while the second
column
is obtained from the susceptibility of the subblocks. $L_1 - L_2$ gives the
pair of lattices, which is used to calculate the ratio of the exponents.
}
\label{gamnu}
\begin{tabular}{ccc}
lattice& \multicolumn{2}{c}{$\gamma / \nu$} \\ \hline
 $L_1 - L_2$&full lattice & subblocks\\ \hline
$  4 - 8 $ & 2.027(4) & 1.898(2) \\
$  8 - 16$ & 1.996(4) & 1.954(3) \\
$ 16 - 32$ & 1.978(4) & 1.966(3) \\
$ 32 - 64$ & 1.979(5) & 1.974(4) \\
\end{tabular}
\end{table}

%
\begin{table}
\caption{
Expectation values of the
specific heat $C_h$, the second moment correlation length $\xi_{2^{nd}}$, and
the  product of the helicity modulus $\heli$ times the linear size of the
system that are reweighted to the final estimate of the critical coupling
$\coup_c = 0.45420$. The errors are calulated from a Jackknife analysis.
}
\label{critres}
\begin{tabular}{llll}
 $L$& $C_h$     &  $\xi_{2^{nd}}$ & $\heli \cdot L$  \\ \hline
  4 & 6.561(27) &  \phantom{1}2.3112(34)    &     1.090(2) \\
  8 & 8.877(33) &  \phantom{1}4.6856(68)    &     1.091(4) \\
 16 &10.704(72) &  \phantom{1}9.4639(142)   &     1.12(1)  \\
 32 &12.564(63) &  19.003(34)               &     1.13(2)  \\
 64 &14.406(102)&  38.25(69)                &        -     \\
\end{tabular}
\end{table}

%
\begin{table}
\caption{
Comparison of critical properties determined from various methods. The results
given for the simulations of the model in the high temperature phase
are obtained from fits according to the simple scaling ansatz with the coupling
$K$ as parameter. For $\gamma$ and $\nu$ from Ref.\protect\cite{ferrer73a} we
took the
estimates of the fcc lattice, since the errors are smaller than those obtained
from the sc lattice.
}
\label{res}
\begin{tabular}{lcccccc}
Method              & Ref.             & $K_c$ & $\gamma$ & $\nu$ & $\eta$ &
$U_L$  \\ \hline
Phenomenological RG & this work        & 0.45420(2) & - & 0.662(7) & 0.024(6) &
0.589(2) \\
High temperature MC & this work        & 0.454170(7)&1.324(1)& -   &    -     &
  -      \\
Phenomenological RG & \cite{janke90a}  & 0.4542(1) &- & 0.670(2)& $\approx$0.02
&0.586(1)\\
High temperature MC & \cite{janke90a}  & 0.45408(8)&1.316(5)& - &   -     &   -
       \\
Phenomenological RG & \cite{we90a}     &   -  &   -  &  $\approx$ 0.67& -
&0.590(5) \\
High temperature MC & \cite{we90a}     & 0.45421(8) & 1.327(8) &   - & - & - \\
$\epsilon$-expansion& \cite{guillou85a}&    - & 1.315(7)& 0.671(5)&0.040(3)&
-\\
$\epsilon$-expansion& \cite{brezin85a} &    -&-&-&-& 0.552 \\
Resummed perturbation
series              & \cite{guillou80} &  - & 1.3160(25)& 0.669(2) & 0.033(4) &
 - \\
High temperature series& \cite{ferrer73a}&0.4539(12) & 1.323(15) & 0.670(7)&
0.028(5)& - \\
High temperature series& \cite{Adler}&0.45414(7) & 1.325 & 0.673 & 0.030& - \\
High temperature series& \cite{butera}&0.45406(5)& 1.315(9)&0.68(1)&-&- \\
Experiment $^4$He         & \cite{ahlers} &  - & - & 0.6705(6)&- &-\\
\end{tabular}
\end{table}


\newpage
\begin{thebibliography}{32}

\bibitem{guillou80}
      J.C. Le Guillou and J. Zinn-Justin, Phys. Rev. B {\bf 21}, 3976 (1980).

\bibitem{wilson72}
       K.G. Wilson and M.E. Fisher, Phys. Rev. Lett. {\bf 28}, 240 (1972).

\bibitem{guillou85a}
      J.C. Le Guillou and J. Zinn-Justin, J. Phys. Lett.  (Paris) {\bf 46},
       L137 (1985).

\bibitem{dombgreen}
     {\it  Phase Transitions and Critical Phenomena} vol. 3 , C. Domb and
     M.S. Green eds. (Academic Press, London, 1974).

\bibitem{ferrenberg91a}
      A.M. Ferrenberg and D.P. Landau, Phys. Rev. B {\bf 44}, 5081 (1991).

\bibitem{baillie92a}
      C.F. Baillie, R. Gupta, K.A. Hawick and G.S. Pawley, Phys. Rev.
      B {\bf 45}, 10438 (1992).

\bibitem{swend+sokal}
      For reviews see, e.g.:\newline
      A.D. Sokal, Nucl. Phys. B (Proc. Suppl.) {\bf 20}, 55 (1990);\newline
      R.H. Swendsen, J.S. Wang and A.M. Ferrenberg,
      in {\it The Monte Carlo Method in Condensed Matter Physics},
      ed. K. Binder (Spinger-Verlag, Berlin-Heidelberg, 1992).

\bibitem{we90a}
      M. Hasenbusch and S. Meyer, Phys. Lett. B {\bf 241},  238 (1990).

\bibitem{janke90a}
      W. Janke, Phys. Lett. A {\bf 148},  306 (1990).

\bibitem{Wang}
      R.\,H.\,Swendsen and
      J.-S.\,Wang, Phys.\ Rev.\ Letters {\bf 58}, 86 (1987).

\bibitem{wolff89a}
      U. Wolff, Phys. Rev. Lett. {\bf 62},  361 (1989);\newline
      U. Wolff, Nucl. Phys. {\bf B322},  759 (1989).

\bibitem{frohlich76}
      J. Fr\"ohlich,  B. Simon and T. Spencer, Commun. Math. Phys. {\bf 50}, 79
      (1976).

\bibitem{frohlich82}
      J. Fr\"ohlich and T. Spencer, Commun. Math. Phys. {\bf 83}, 411 (1982).

\bibitem{kohring86}
      G. Kohring, R.E. Shrock and P. Wills, Phys. Rev. Lett. {\bf 57},  1358
      (1986).

\bibitem{wolff90a}
      U. Wolff, Nucl. Phys. B {\bf 334}, 581 (1990).

\bibitem{hasenbusch90a}
      M. Hasenbusch, Nucl. Phys. {\bf B333},  581 (1990).

\bibitem{sokal92a}
      F. Cooper, B. Freedman and D. Preston, Nucl. Phys B {\bf 210}, 210
      (1982);\newline
      R.G. Edwards, S.J. Ferreira, J. Goodman and A.D. Sokal, Nucl. Phys. B
      {\bf 380}, 621 (1992).

\bibitem{janke92a}
      C. Holm and W. Janke, HLRZ preprint 77/92 J{\"u}lich (1992).

\bibitem{fisher73a}
      M.E. Fisher, M.N. Barber and
      D. Jasnow, Phys. Rev. A  {\bf 8}, 1111 (1973).

\bibitem{liteitel}
      Y.-H. Li and S. Teitel, Phys. Rev. B {\bf 40}, 9122 (1989).


\bibitem{siam}
      R.G. Miller, Biometrica {\bf 61}, 1 (1974); \newline
      B. Efron, {\it The Jackknife, the Bootstrap and other Resampling Plans}
      (SIAM, Philadelphia, PA, 1982).

\bibitem{wegner}
     F.J. Wegner, Phys. Rev. B{\bf 5}, 4529 (1972)  

\bibitem{esserdohm}
     A. Esser and V. Dohm, in {\it The $18^{th}$ IUPAP International Conference
     on Statistical Physics} p.159, ed. W. Loose (TU Berlin, Germany, 1992).



\bibitem{binder81b}
      K. Binder, Phys.Rev.Lett {\bf 47}, 693 (1981) ; \newline
      K. Binder, Z. Phys. B - Condensed Matter {\bf 43}, 119 (1981); \newline
      K. Binder, M. Nauenberg, V. Privman and  A.P. Young, Phys. Rev. B
      {\bf 31}, 1498 (1985).

\bibitem{ferrer73a}
      M. Ferer, M.A. Moore and W. Wortis, Phys. Rev. B {\bf 8}, 5205 (1973).

\bibitem{Adler}
      J. Adler, C. Holm and W. Janke, preprint FUB-HEP 18/92 , HLRZ 76/92
      Berlin/J\"ulich (1992).

\bibitem{butera} P. Butera, M. Comi and A.J. Guttmann, preprint IFUM 444/FT
      Milano (1992).

\bibitem{luescher}
     M. L\"uscher and P. Weisz, Nucl. Phys. B {\bf 300}, 325 (1988).

\bibitem{privat}
      W. Janke, private communication.

\bibitem{ahlers} L.S. Goldner and G. Ahlers, Phys. Rev. B {\bf 45}, 13129
      (1992).

\bibitem{brezin85a}
     E. Br\'{e}zin and J. Zinn-Justin, Nucl. Phys. B {\bf 257}, 867 (1985).

\end{thebibliography}
\end{document}